\documentclass[11pt,a4paper]{article}
\usepackage{jheppub}
\usepackage{pifont}
\usepackage{natbib}
\usepackage{amsmath}
\usepackage{graphicx}
\usepackage{enumerate}
\usepackage{epstopdf}
\usepackage{subfig}
\usepackage{dsfont}
\usepackage{amssymb}
\usepackage{amsthm}
\usepackage{mathrsfs}
\usepackage{cases}
\title{Bound states in the continuum of fractional Schr\"{o}dinger equation in the Earth's gravitational field and their effects in the presence of a minimal length: applications to distinguish ultralight particles}
\author[a]{Zhang Xiao,}
\author[b]{Yang Bo,}
\author[c]{Wei Chaozhen,}
\author[a]{Luo Maokang}

\affiliation[a]{Department of Mathematics, \\Sichuan University 610065, Chengdu, Sichuan, China}
\affiliation[b]{Department of Mathematics, \\Yunnan Normal University 650504, Kunming, Yunnan, China}
\affiliation[c]{Department of Mathematics, \\State University of New York at Buffalo, Buffalo, NY, USA}

\emailAdd{destronstain@126.com}
\emailAdd{boyang1012@163.com}
\emailAdd{makaluo@scu.edu.cn}

\abstract{In this paper, the influence of the fractional dimensions of the L\'{e}vy path under the Earth's gravitational field is studied, and the phase transitions of energy and wave functions are obtained: the energy changes from discrete to continuous and wave functions change from non-degenerate to degenerate when dimension of L\'{e}vy path becomes from integer to non-integer. By analyzing the phase transitions, we solve two popular problems. First, we find an exotic way to produce the bound states in the continuum (BICs), our approach only needs a simple potential, and does not depend on interactions between particles. Second, we address the continuity of the energy will become strong when the mass of the particle becomes small. By deeply analyze, it can provide a way to distinguish ultralight particles from others types in the Earth's gravitational field, and five popular particles are discussed. In addition, we obtain analytical expressions for the wave functions and energy in the Earth's gravitational field in the circumstance of a fractional fractal dimensional L\'{e}vy path. Moreover, to consider the influence of the minimal length, we analyze the phase transitions and the BICs in the presence of the minimal length. We find the phenomenon energy shift do not exist, which is a common phenomenon in the presence of the minimal length, and hence such above phenomena can still be found. Finally, relations between our results and existing results are discussed.}

\begin{document}

\maketitle
\flushbottom
\section{Introduction}
Over the past few decades, fractional quantum quantization methods has been put forward by Laskin and us \cite{3,6,7}, which is a natural generalization of standard quantum quantization methods that arises when the Brownian trajectories are replaced by L\'{e}vy flights or systems satisfy the fractional corresponding relation. Recently, some of these theoretical results were realized in Refs. \cite{65,68}. These lead to fractional quantum mechanics, which can describe non-Markovian evolution with a memory effect and nonlocal quantum phenomena \cite{35}, and enables a path for studying route to fractional dimensions \cite{2,6,7}. A large number of studies have been reported in many aspects, like fractional Schr\"{o}dinger equations \cite{3,6,7,32,35,36,37,52}, profile decompositions with angularly regular data \cite{40}, optimal controls \cite{39}, transmission through locally periodic potentials \cite{38}, polariton condensates \cite{66}, long-range dynamics and quantum transport \cite{67}, and various other potential fields \cite{1,33,34,46,47}; however, fractional quantum mechanics can also describe phenomena that classical quantum mechanics is unable to do, such as one-dimensional L\'{e}vy crystals \cite{28,29}, exotic BICs \cite{2} and propagation dynamics of a light beam \cite{65}.\par
In this paper, we obtain two important results on fractional quantum systems and ultralight particles by studying the effects of the fractional dimensions of the L\'{e}vy path which has been realized through a designed experiment \cite{68}. Since these two problems usually need to be considered in a simpler and easily realized potential, we choose to study a popular potential, the Earth's gravitational field, which has been studied for a long time and can be easily realized. No articles reporting the effects of the fractional dimensions on the Earth's gravitational field have been published; however, in this paper, we find that the effects would cause some important results for providing the answer of the problems: exotic method to produce BICs phenomenon and characteristic of ultralight particles to distinguish them from others, it is straightforward to expect effects in a gravitational field.\par
In our approach, we solve the two problems in the following four steps:
\begin{enumerate}[(a)]\setlength{\itemsep}{0pt}
\item
Starting from the fractional corresponding operator $T_\alpha$ \cite{5}, which relates to the momentum of the fractional path \cite{3}, then using our fractional quantum quantization method, we establish the general fractional Schr\"{o}dinger equation [Eq. \eqref{eq:FSE-F}] to describe the problems.
\item
The fractional quantization method allows us to choose the most convenient fractional derivative to describe a particular problem depending on its specific characteristics, rather than only using the Riesz fractional derivative operator \cite{1,6,7,34,36}. Moreover, this convenient choice does not influence properties of the problem, such as wave functions and energy \cite{3}. Thus, by analyzing the different effects of different fractional derivatives to deal with problems (see Appendix \ref{app-d} for more details), we choose the Riemann-Liouville fractional derivative.
\item
We find that the wave functions are determined by the function with the form \eqref{eq:aux0}, which includes the parameter $2\alpha$. Then, we prove that the linear combination of the function are wave functions. Thus, we obtain analytic expressions of energy and wave functions in the circumstance of a fractional dimensional L\'{e}vy path, which contains the effects of the fractional dimension.
\item
The results indicate that a fractional fractal dimension will cause a phase transition: the energy changes from discrete to continuous and wave functions change from non-degenerate to degenerate when dimension of L\'{e}vy path becomes from integer to non-integer, these make all bound state energy levels of the particle continuous, whose degeneracies are strongly related to the dimension of the L\'{e}vy path. These phenomena are remarkably different from those in the classical case, where the particle's bound state energies in the Earth's gravitational field are discrete and have no degeneracy. However, when the dimension of the L\'{e}vy path becomes an integer, our results coincide with the classical results.
\end{enumerate}\par
To the first problem, the phase transition provides an exotic method for causing the BICs phenomenon and is different from methods previously considered by many authors \cite{54,55,59,61,62}. Previous BICs can only be realized in two types of ways: one type of BICs is fragile and realized in tailored potentials without interactions between particles \cite{11,27,60}; the other type is realized through interactions between particles \cite{63,64}. However, our BICs can be realized in a simple potential without interactions between particles, and not in specially tailored potential and/or through the interactions between particles. To the best of our knowledge, it is also the first BICs found in the Earth's gravitational field. We clarify that this phenomenon can be considered as a characteristic phenomenon in circumstance of fractional dimensional L\'{e}vy path, and it provides a criterion to determine in advance whether an unknown system can be described by fractional derivatives. There is evidence that the BICs phenomenon can be observed in certain experiments \cite{69}, and the above phenomenon may also be used as a sufficient condition to verify successful preparation of a fractional quantum mechanical system in experiments.\par
The system of a particle moving in the Earth's gravitational field has been studied for a long time in classical quantum mechanics; however, the effects caused by the fractional dimensions have yet to be reported in the literature. Although some authors have considered a particle moving in the linear field with a L\'{e}vy path of fractional dimensions \cite{1}, they used the method in classical quantum mechanics and hence obtained similar results as those in classical quantum mechanics: the energy of a moving particle is discrete and non-degenerate. By our analysis, their results turn out to be incomplete because they did not include the effects of fractional dimensions. The details will be discussed in Section \ref{sec:2a}.\par
For the second problem, by deeply analyzing the relations between phase transition and mass of particle, we find an essential property of the energy of ultralight particles that can distinguish them from other particles. In recent years, ultralight particles have played an important role in the field of particle physics and dark matter \cite{70,74}. We found that the continuity of energy becomes strong when (1) the dimension of the L\'{e}vy path changes from integer to non-integer and/or (2) the mass of the particle becomes small. Conversely, in these two cases, the energy will be asymptotically close to that in classical quantum mechanics; see \cite{9,10,11} and experimental results in Refs. \cite{12,14}. We compare the energies of five different particles: the Z boson, tau, muon, electron and electron neutrino. The results illustrate that we can easily utilize this property to distinguish ultralight particles from other particles. Since the accurate observation of the energy levels of particles in the Earth's field has been realized in Ref. \cite{12}, this property may provide some improvements in the detection of ultralight particles.\par

Moreover, to consider the effects of the minimal length on the the phase transitions and the BICs, whose existence is predicted in many theories including string theory, loop quantum gravity and black hole physics \cite{17,19,50,51,72,75,76}.

%%many papers have reported that the minimal measurable length can exist \cite{15,16,17} and cause influence to the quantum gravity \cite{41,42,43,44,48,49}. The influence can be considered as a universality of quantum gravity corrections \cite{17}.
We propose to study the influence of the minimal length under the Earth's gravitational field in the fractional dimensional L\'{e}vy path. We build the fractional Schr\"{o}dinger equation in the presence of the minimal length, and the analytic expressions of the energy and corresponding wave functions are provided. We find a common phenomenon---energy shift \cite{48,49}---in the presence of the minimal length does not occur. The results show that the minimal length does not change the structure of the wave functions and energy levels in the fractional dimension, i.e., the phase transitions and the BICs still exist. To the best of our knowledge, this is the first time that such BICs have been reported in the presence of the minimal length.

%
%Then
%
%
%
%
%to find the phase transitions of energy and wave functions.
%
%
%by using the generalized de Moivre's theorem \cite{2} and the method for solving a specific kind of differential equations (see Appendix \ref{app-a} for the details).  \par  \par
% %%In fundamental principles, the studies includes . And in applications, fractional quantum mechanics has a very wide applications including \par

%In this paper, we study a nonrelativistic particle subjected to the Earth's gravitational field \par
%
%
%
%
%
%
%And the circumstance has been realized \cite{65,68}. We introduce
%
%
%

%%all the energies of bound states are continues and the minimum measurable do not cause the energy shift, which are different from the results in classical situation \cite{48,49}.
%%
\section{Earth's gravitational field}\label{sec:2}
\subsection{Earth's gravitational field in the circumstance of a fractional fractal dimensional L\'{e}vy path}
As mentioned above, we first introduce the equation that is proved in \cite{3}
\begin{equation}\label{eq:begin}
<p^\alpha>=\int_{-\infty}^{+\infty}\varphi^*(x,t)[(-i\hbar)^\alpha T_\alpha]\varphi(x,t)\mathrm{d}x,
\end{equation}
where $2\alpha\in(1,2]$ is the dimension of the L\'{e}vy path \cite{2,3}, $T_{2\alpha}$ is a fractional corresponding operator defined in Refs. \cite{2,3}, $<p>$ is the expectation of the momentum of quantum system, and $\varphi$ is a wave function. Eq. \eqref{eq:begin} indicates the form of the momentum of quantum system in the circumstance of a fractional L\'{e}vy path, thus it can provide the fractional corresponding relation \cite{3}:
\begin{equation}\label{eq:corresponding}
p^{2\alpha}\rightarrow  (-i\hbar)^{2\alpha} T_{2\alpha},
\end{equation}
This relation is a natural generalization of the classical corresponding relation $p^2\rightarrow  -\hbar^2 \frac{\partial^2}{\partial x^2}$. It not only indicates the momentum of a quantum system in the circumstance of a fractional L\'{e}vy path but also establishes the connection between two different fields: quantum mechanics and fractional calculus \cite{3,6}. When $\alpha=1$, Eq. \eqref{eq:corresponding} becomes the classical corresponding relation \cite{3}.\par
Using the fractional corresponding relation [Eq. \eqref{eq:corresponding}] in the Hamiltonian of the form \cite{6,7} %%forºÏÊÊÂð£¿4
\begin{equation}\label{eq:ham}
H=D_{2\alpha}p^{2\alpha}+V,
\end{equation}
we can build the general fractional Schr\"{o}dinger equation:
\begin{equation}\label{eq:FSE-F}
i\hbar\frac{\partial\varphi}{\partial t}=D_{2\alpha} (-i\hbar)^{2\alpha} T_{2\alpha}\varphi+V\varphi,
\end{equation}
where $D_{2\alpha}$ has dimensions of $erg^{1-2\alpha}\times cm^{2\alpha}\times sec^{-2\alpha}$ \cite{6,7}, and $T_{2\alpha}$ is defined as follows:
A linear operator $T_\alpha$ of order $\alpha\in (0,+\infty)$ with respect to $x$ is called a fractional corresponding operator if it satisfies the following conditions:
\begin{enumerate}[(a)]\setlength{\itemsep}{0pt}
\item \label{condition a}
\begin{equation*}
T_\alpha\delta(\frac{x+y}{c})=\frac{i^\alpha}{2\pi c^\alpha}\int_{-\infty}^{+\infty}k^\alpha \exp[{ik(\frac{x+y}{c})}]\mathrm{d}k,
\end{equation*}
for any $x,y\in\mathds{R}$, where $c\neq0$ is a constant.
\item
\begin{equation*}
T_\alpha[g(x)l(y)]=l(y)T_\alpha g(x),
\end{equation*}
for any $g(x)$ and $l(y)$ being continuous functions.
\item
\begin{equation*}
T_\alpha\int_{-\infty}^{+\infty}f(x,t)\mathrm{d}t=\int_{-\infty}^{+\infty}T_\alpha f(x,t)\mathrm{d}t,
\end{equation*}
for any $f(x,t)\in L^2(\mathds{R}^2)$.
\item
When $\alpha$ approaches $n$, where $n$ is a positive integer, $T_n$ is a classical derivative operator.
\end{enumerate}\par
Although Eq. \eqref{eq:FSE-F} contains the abstract operator $T_{2\alpha}$, there is no significant difficulty in analyzing problems since the operator satisfies an important property, i.e., the five popular fractional derivative operators coincide with the fractional corresponding operator: the R-L fractional derivative operator, the Gr\"{u}nwald-Letnikov fractional derivative operator, the Caputo fractional derivative operator, the Riesz fractional derivative operator, and the fractional derivative operator based on generalized functions. We will be able to choose a comfortable operator to address the Earth's gravitational field. If choosing the Riesz fractional derivative operator, which can easily address calculations related to Fourier transforms, and following the classical method (transform the Schr\"{o}dinger equation from the space representation to the momentum representation), which is used to solve the Earth's gravitational field in the classical situation, the results \cite{1} obtained did not contain the effect caused by the fractional dimension. Their results showed that every energy level $E_n$ has only one corresponding wave function $\varphi_n$, which is similar to the case in classical quantum mechanics.

Applying this method to classical quantum mechanics is actually appropriate since the wave functions in classical quantum mechanics are nondegenerate. However, in fractional cases, fractional dimensions would cause the wave functions to be degenerate \cite{2}. Thus, using this method they would only obtain parts of the solutions. This is because $\varphi(x)$ in the position representation and $\psi(p)$ in the momentum representation do not always have a one-to-one correspondence \cite{8}, i.e., the Fourier transforms of different $\varphi(x)$ have the same form of $\psi(p)$. Thus, the wave functions obtained by Eq. \eqref{eq:FE} with the Riesz fractional derivative in the momentum representation are parts of Eq.\eqref{eq:g-solution}. \par
In our approach, we analyze Eq. \eqref{eq:FSE-F} using the fractional quantization method which allows us to choose the most convenient fractional derivative to describe a particular problem depending on its specific characteristics. By analyzing the different effects of different fractional derivatives to deal with the problem, we choose the Riemann-Liouville fractional derivative.\par
Let us consider the gravitational field defined by
\begin{equation}\label{eq:gravitational field}
V(z)=
\begin{cases}
+\infty & \text{$z<0$}\\
mgz & \text{$z\geq 0$}
\end{cases}
,
\end{equation}
which can form a bound potential well. In the region $z\geq 0$, the general fractional Schr\"{o}dinger equation can be written as
\begin{equation}\label{eq:FE}
D_{2\alpha}(-i\hbar)^{2\alpha}{}_{-\infty}^R\!D_z^{2\alpha}\varphi(z)+mgz\varphi(z)=E\varphi(z),
\end{equation}
where
\begin{equation*}
{}_{-\infty}^R\!D_z^{2\alpha}\varphi(z)=\frac{1}{\Gamma(m-2\alpha)}\frac{\partial^m}{\partial z^m}\int_a^z(z-u)^{m-2\alpha-1}\varphi(u)\mathrm{d}u,
\end{equation*}
is the R-L fractional derivative operator \cite{4,5}
\subsection{Phase transitions of wave functions and energy}\label{sec:2a}

%Recently, in Ref. \cite{1}, authors discussed Eq. \eqref{eq:GFSE} with Riesz fractional derivative operator in the momentum representation and then solved it. \\\\

Since it is difficult to obtain analytic solutions of Eq. \eqref{eq:FE} containing the effect of the fractional dimension directly, we introduce an auxiliary function $y_\nu$:
\begin{equation}\label{eq:aux0}
y_\nu(z)=\varepsilon_\nu\int_0^{+\infty}\exp[\varepsilon_\nu zt-\frac{t^{2\alpha+1}}{K(2\alpha+1)}]\mathrm{d}t,
\end{equation}
where $\varepsilon_\nu^{2\alpha+1}=1$ indicates the structure of wave functions in the fractional path, $2\alpha$ is the dimension of the L\'{e}vy path, and $K=\{[{D_{2\alpha}(-i\hbar)^{2\alpha}}]/({-mg})\}^{2\alpha}$ is a parameter determined by the particle's mass. Then, we will prove that the linear combination of $y_\nu(z)$ can form the wave functions.\par

Let $\xi=({E-mgz})/[{D_{2\alpha}(-i\hbar)^{2\alpha}}]$ and $\psi[\xi(x)]=\varphi(x)$ using the fractional chain rule \cite{25}, and substituting the parameter $K$, Eq. \eqref{eq:FE} becomes
\begin{equation}\label{eq:GFSE-2}
{}_{-\infty}^R\!D_\xi^{2\alpha}\psi(\xi)-K\xi\psi(\xi)=0.
\end{equation}\par
Assuming $\{C_\nu\}$ is a sequence of arbitrary coefficients subject to $\sum C_\nu=0$, for an arbitrary energy level $E$, we can find that the solutions of Eq. \eqref{eq:GFSE-2} (see Appendix \ref{app-a} for a description of this method and related mathematical deductions) are
\begin{equation}\label{eq:g-solution}
\psi(\xi)=\sum C_\nu y_\nu(\xi).
\end{equation}\par
The known constraint conditions cannot determine all the undetermined parameters $C_\nu$ as we shall discuss below. We can arbitrarily adjust the value of the free undetermined parameters to make Eq. \eqref{eq:g-solution} satisfy Eq. \eqref{eq:bound condition}. Thus, when $\alpha\neq1$, it is easy to make Eq. \eqref{eq:g-solution} satisfy the bound conditions:
\begin{gather}\label{eq:bound condition}
\left\{
  \begin{array}{ll}
\psi[\frac{mgz-E}{D_{2\alpha}(-i\hbar)^{2\alpha}}]|_{z=0}=0\\
\psi[\frac{mgz-E}{D_{2\alpha}(-i\hbar)^{2\alpha}}]|_{z=+\infty}=0
\end{array}
\right.
.
\end{gather}\par
Next, we will discuss phase transitions of the energy and degeneracy.\par
By the generalized de Moivre's theorem, we have three cases for $\alpha$. \par
Case I: $\alpha\neq 1$ is a rational number. Since $\alpha=n/m\in(0.5,1)$, we can show that the minimum numerator of $(2n+m)/m$ is $5$ (see Appendix \ref{app-b} for a strict mathematical proof). Thus, the equation $\varepsilon_\nu^{2\alpha+1}=1$ will have at least $5$ roots and hence $5$ corresponding $y_\nu$. Then, there are $5$ undetermined parameters $C_\nu$ in Eq. \eqref{eq:g-solution}, which require at least $5$ conditions to be determined. However, there are only $4$ conditions, i.e., $\sum C_\nu=0$, Eq. \eqref{eq:bound condition}, and normalization conditions. Thus, the energy $E$ is arbitrary. The energy levels are continuous because of the free parameters. \par
Case II: $\alpha$ is an irrational number. Then, there are countably infinite undetermined parameters $C_\nu$ in Eq. \eqref{eq:g-solution}. We need at least countably infinite conditions to determine all the parameters. However, we only have a finite number of conditions, which are not sufficient to determine all the parameters $C_\nu$; thus, the energy $E$ is arbitrary. Again, the energy levels are continuous. \par
Case III: $\alpha=1$. This is the classical case. We can know that there are $3$ undetermined parameters in Eq. \eqref{eq:g-solution}, and we have sufficient conditions to determine all the unknown parameters. Thus, the energy $E$ is not arbitrary, which implies that the energy levels are discrete, consistent with the results of classical quantum mechanics \cite{9,10,11}.\par
When $\alpha=1$, the corresponding eigen-states of energy levels have no degeneracy \cite{9,10,11}. However, when $\alpha\in(0.5,1)$, the index $\nu$ is at least $N(\geq 5)$. Since we do not have sufficient constraint conditions to determine all the unknown parameters in Eq. \eqref{eq:g-solution}, some different eigen states of an arbitrary energy $E$ are caused by different values of the undetermined parameters. Hence, the energy $E$ would exhibit degeneracy, and it is strongly related to the dimension $2\alpha$ of the L\'{e}vy path.\par
As a result, when dimension of L\'{e}vy path becomes from integer to non-integer, the energy of the particle changes from discrete to continuous; And the wave functions change from non-degenerate to degenerate which are decided by the dimension $2\alpha$ of the L\'{e}vy path.
\subsection{Bound states in the continuum and ultralight particles}\label{sec:2b}
Before we discuss the exotic BICs, we present results to show how the fractional dimension of the L\'{e}vy path affects the energy.\par
\begin{figure*}[!h]
\centering
\subfloat[$\alpha=0.667$]{\includegraphics[width=2in]{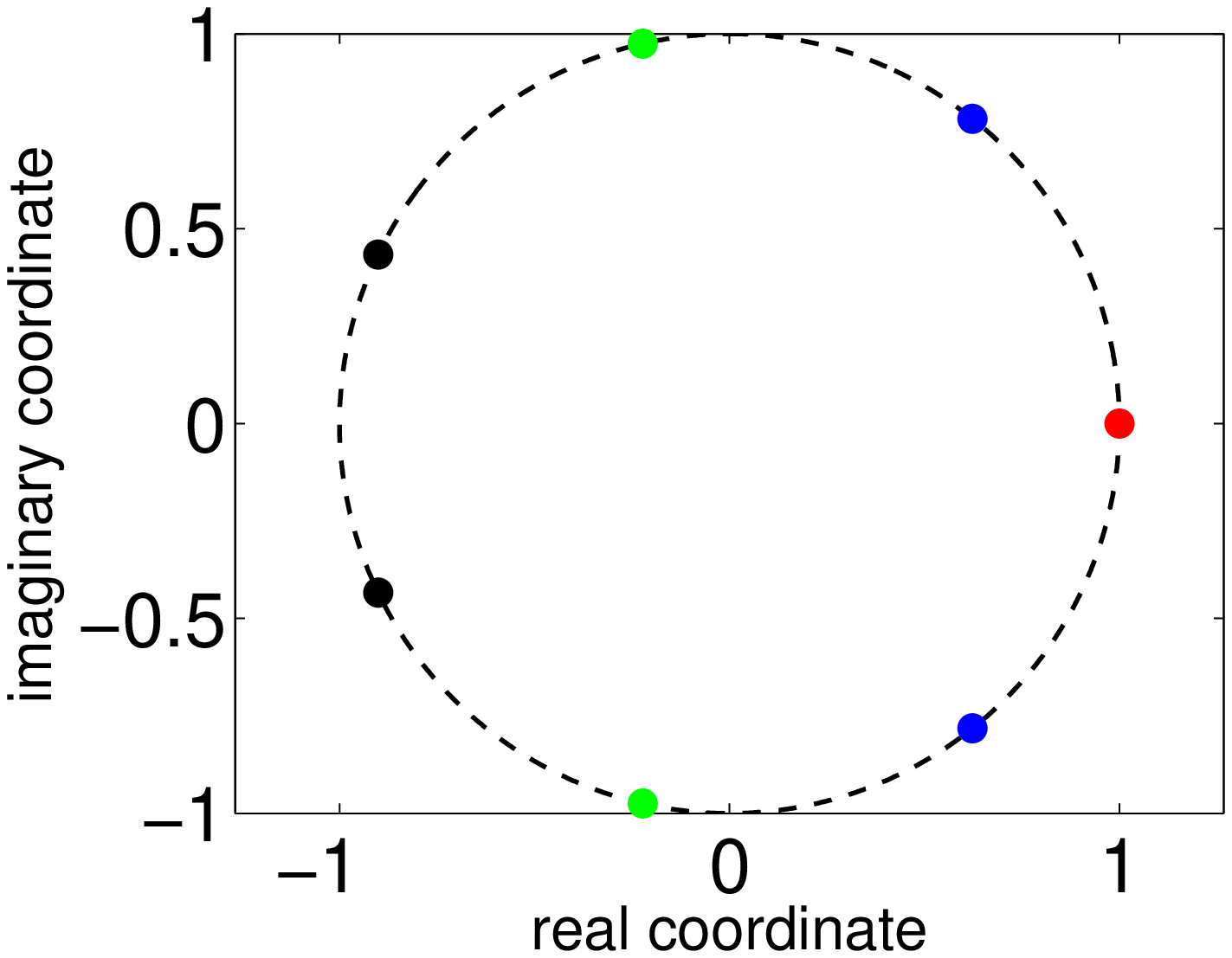}}
\subfloat[$\alpha=0.750$]{\includegraphics[width=2in]{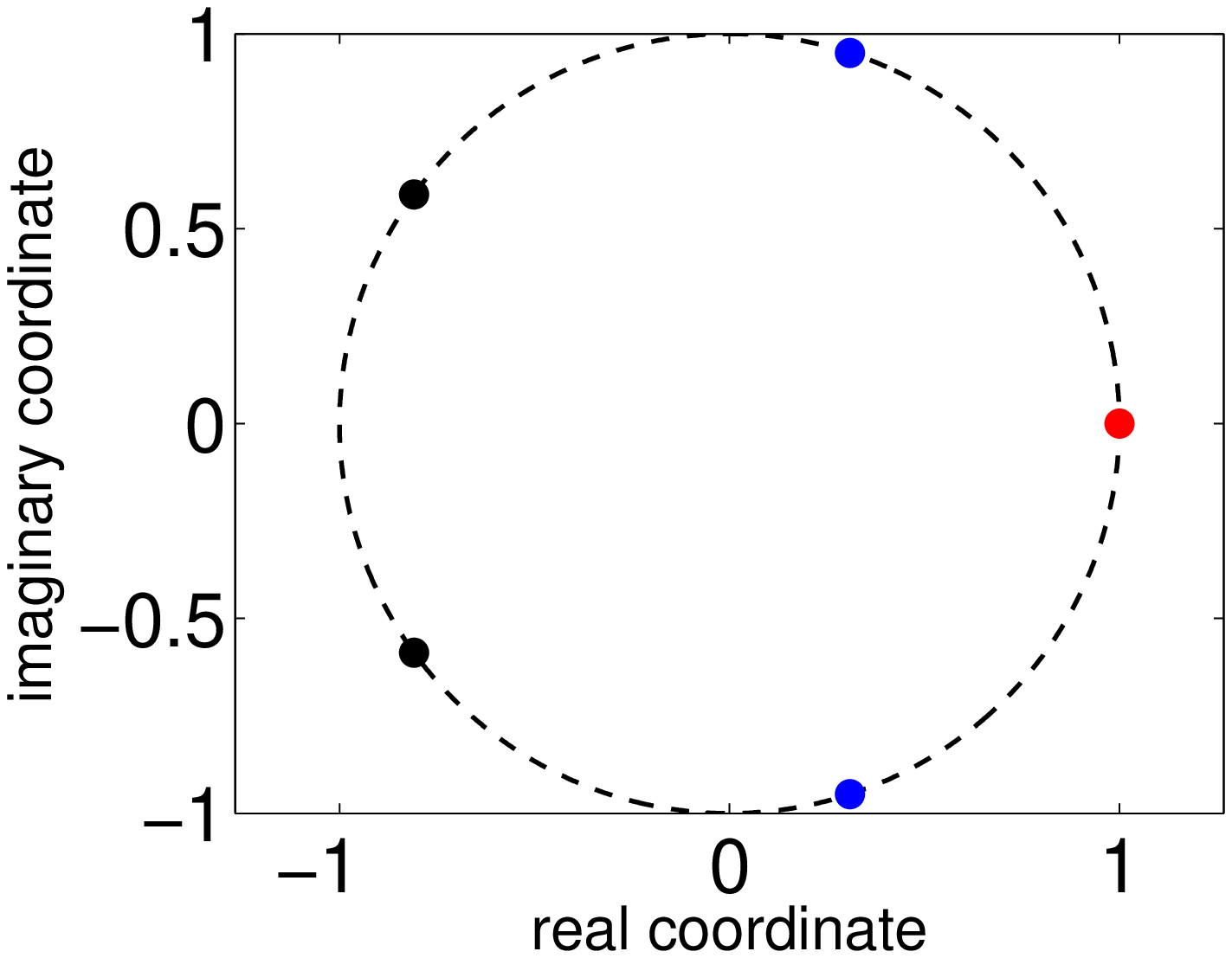}}
\subfloat[$\alpha=0.833$]{\includegraphics[width=2in]{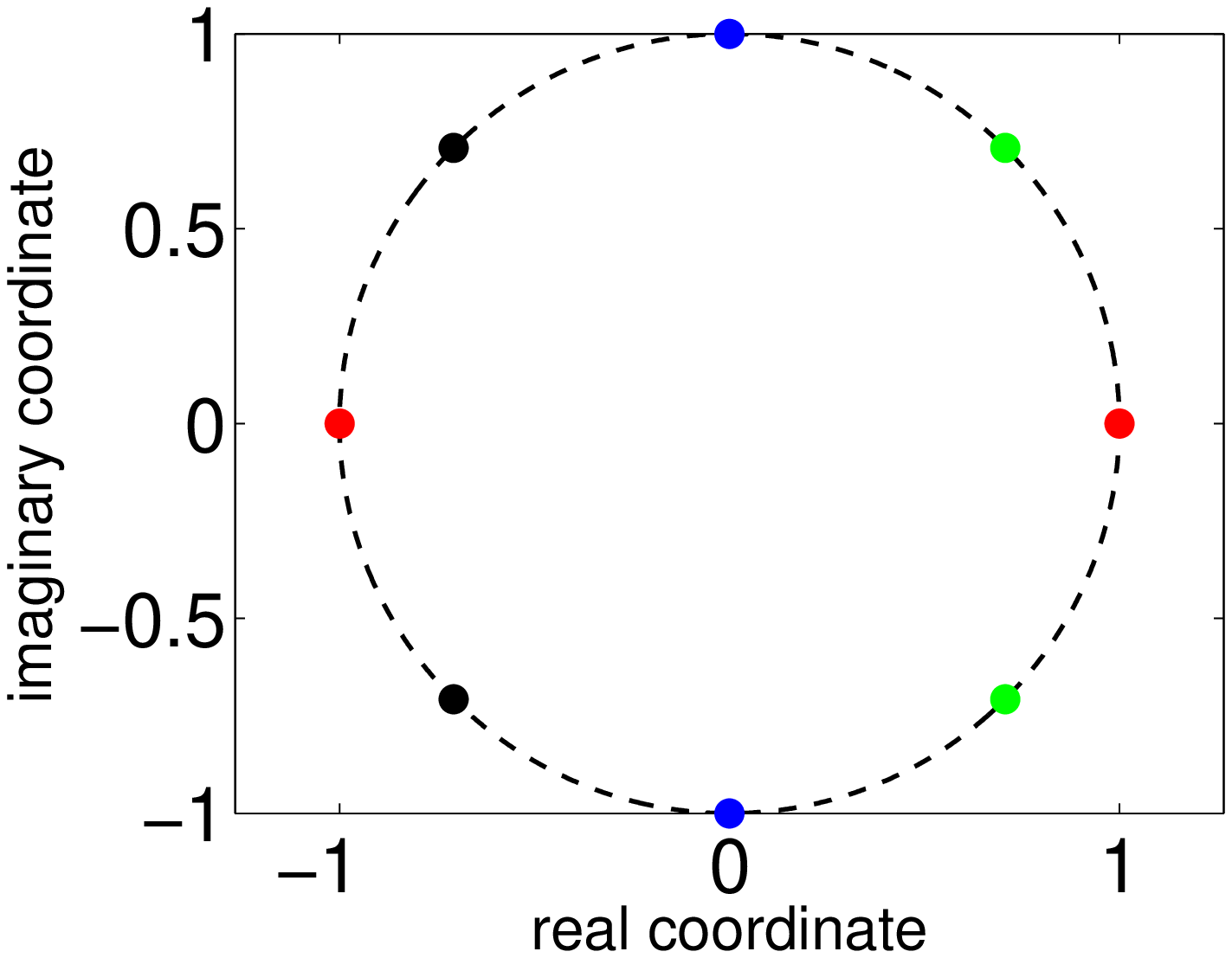}}\\
\subfloat[$\alpha=0.667$]{\includegraphics[width=2in]{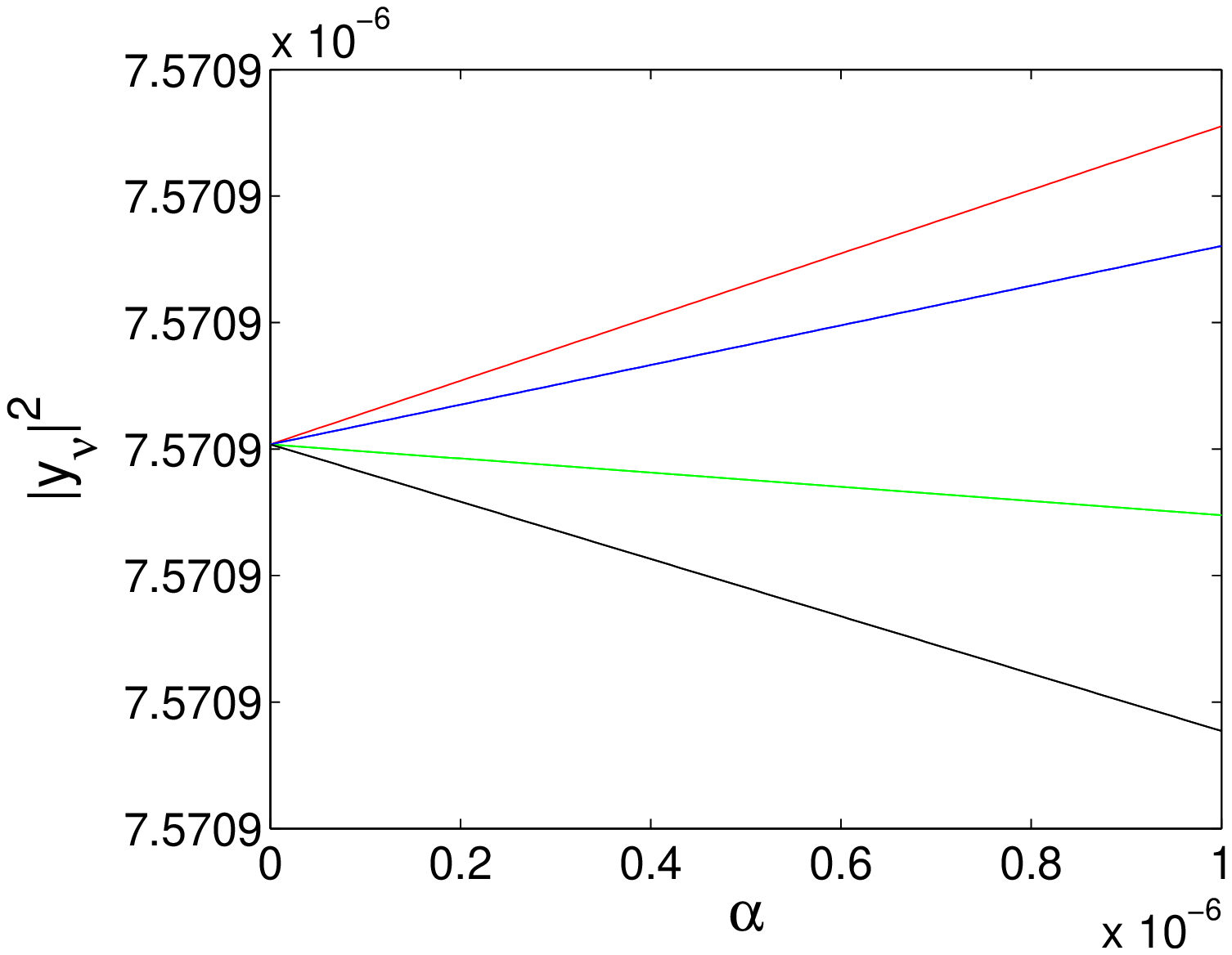}}
\subfloat[$\alpha=0.750$]{\includegraphics[width=2in]{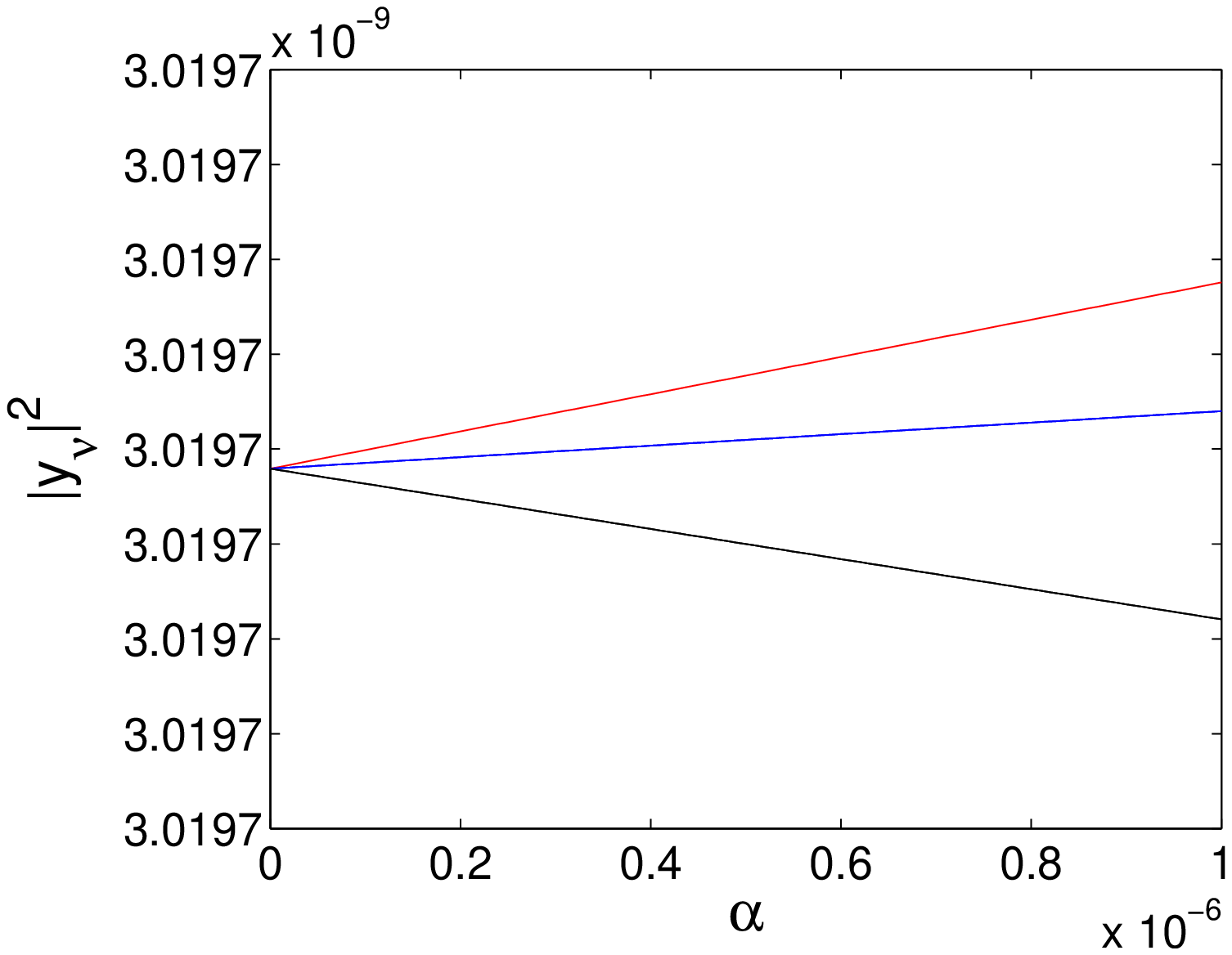}}
\subfloat[$\alpha=0.833$]{\includegraphics[width=2in]{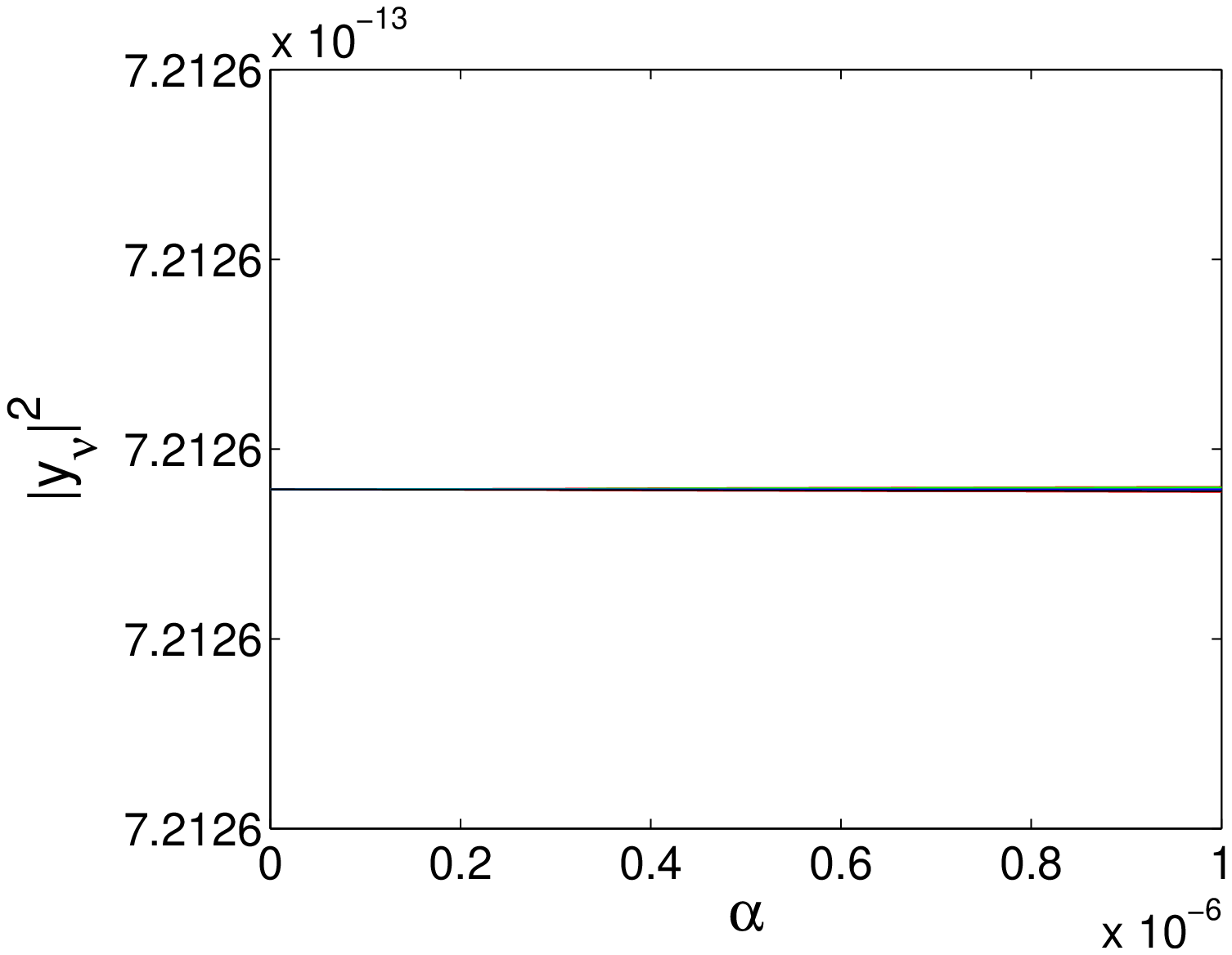}}
\caption{$\varepsilon_\nu$ and its corresponding $|y_\nu|^2$ in different dimensions, where we use the neutron's mass. The number of the different $y_\nu$ which determine energy states are decided by the $\varepsilon_\nu$, and differences between the different $|y_\nu|^2$ determine continuity of the energy decrease quickly when $\alpha$ approach to $1$. These coincides with the results from classical quantum mechanics.}\label{fig:1}
\end{figure*}

The energy states are determined by the number of undetermined parameters $C_\nu$ and constraints and hence by the number of the different $y_\nu$. Thus, first, we discuss the number of the different $y_\nu$, by the generalized de Moivre's theorem, we can calculate the number of the different $y_\nu$ in two different cases: (1) if $\alpha=n/m$, $m,n\in \mathds{Z}$, the number of the different $y_\nu$ is $2n+m$; (2) if $\alpha$ is an irrational number, the number of the different $y_\nu$ is countably infinite. Since $y_\nu$ may has the corresponding conjugate one, using $|y_\nu|^2$ may not clearly show the number of the different $y_\nu$. Thus, we consider another quantity $\varepsilon_\nu$, which is fractional quantum number and in one-to-one correspondence with $y_\nu$. Fig. \ref{fig:1}(a)-(c) illustrates the positions of the fractional quantum number $\varepsilon_\nu$ in the complex plane for the same three specific cases. The colored points in Fig. \ref{fig:1}(a)-(c) and the the corresponding lines in Fig. \ref{fig:1}(d)-(f) with the same color represent the $\varepsilon_\nu$ and the norm of the same solution $y_\nu$, respectively, for different $\alpha$.\par
Second, if the differences between different $y_\nu$ become sufficient small, we can approximately treat different $y_\nu$ in the same manner. Then, the number of free undetermined parameters would decrease; hence, the energy of the system transforms from continuous to discrete. In classical situation(the dimension of the L\'{e}vy path is $2$), we know the energy of the system is discrete; however when the dimension of the L\'{e}vy path approaches from non-integer to $2$, the differences between the different $|y_\nu|^2$ decrease quickly. As we discuss in Section \ref{sec:2a}, the energy in the circumstance of a fractional dimensional L\'{e}vy path is continuous because of the differences between the different $|y_\nu|^2$. That is when the different $|y_\nu|^2$ decrease, the continuity of energy becomes weak, which coincides with the results from classical quantum mechanics \cite{9,10,11}; conversely, the continuity of energy becomes strong when the dimension of the L\'{e}vy path changes from integer to non-integer. Figure \ref{fig:1}(d)-(f) illustrates the different $|y_\nu|^2$ for three different $\alpha$.

\par 
As one of the main conclusions, we provide an exotic BICs. Since the states in the potential field [Eq. \eqref{eq:gravitational field}] are always bounded, the BICs exist. They have the three features simultaneously: all bound state energies of the particles are continuous; particles can always remain in bound states; and they can be realized in a simple potential rather than in specially tailored potentials \cite{11,27,60} and/or by the interactions between particles \cite{63,64}, in contrast to previous methods considered in \cite{27,57,59,61,62,63,64}. Refs. \cite{2,11,27} demonstrated that BICs do not exist in classical quantum mechanics that can not use the classical model to describe; thus, the existence of such BICs is a characteristic phenomenon in the circumstance of the fractional dimensions of the L\'{e}vy path. Thus, we can regard this as a criterion for determining whether the fractional Schr\"{o}dinger equation should be used to describe a quantum system. Moreover, since there is evidence that the BICs phenomenon can be observed in experiment \cite{69}, it may also provide a method to verify the successful preparation of a fractional quantum system.\par
\begin{figure*}[!h]
\centering
\subfloat[$Electron$]{\includegraphics[width=2in]{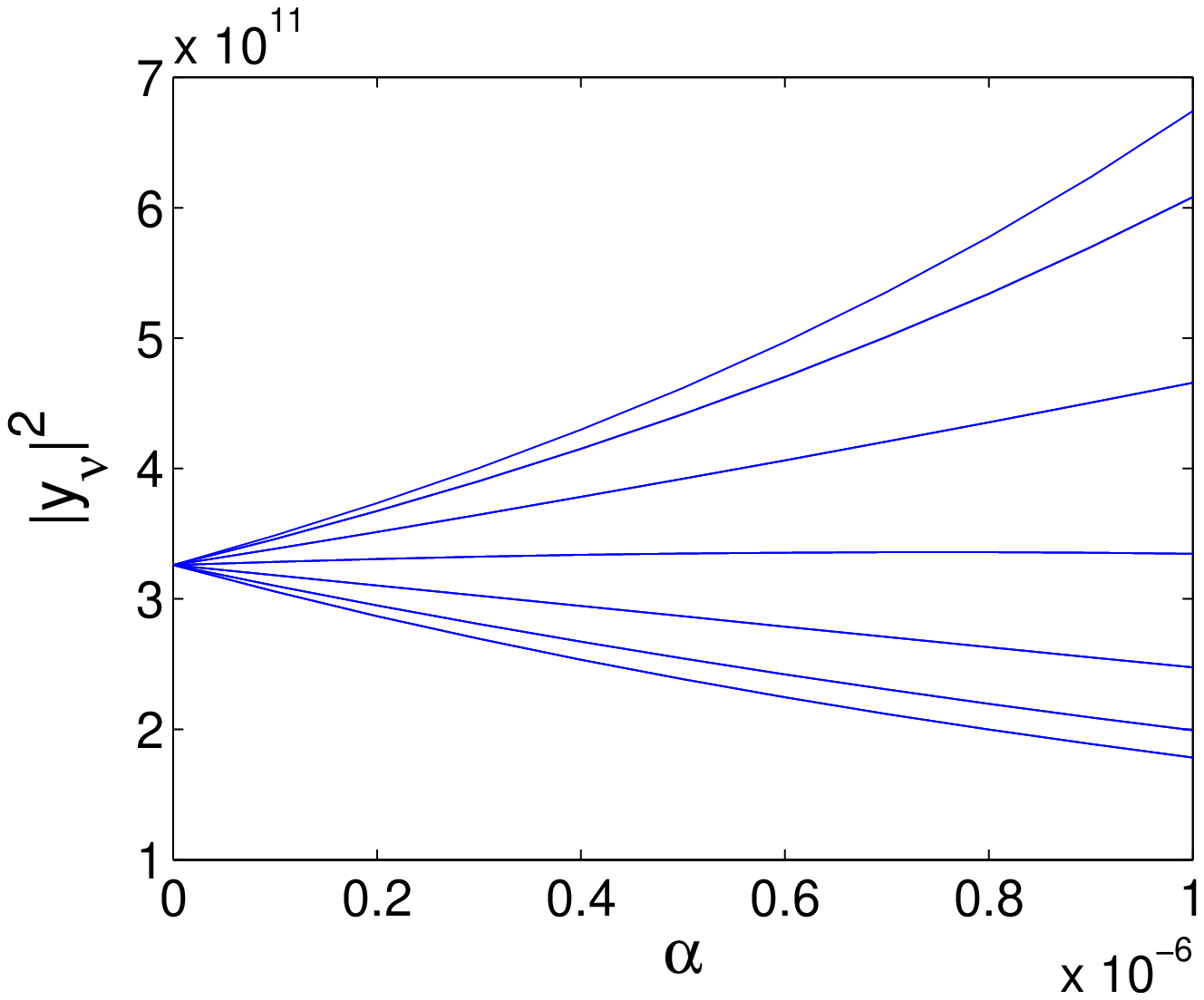}}
\subfloat[$Muon$]{\includegraphics[width=2in]{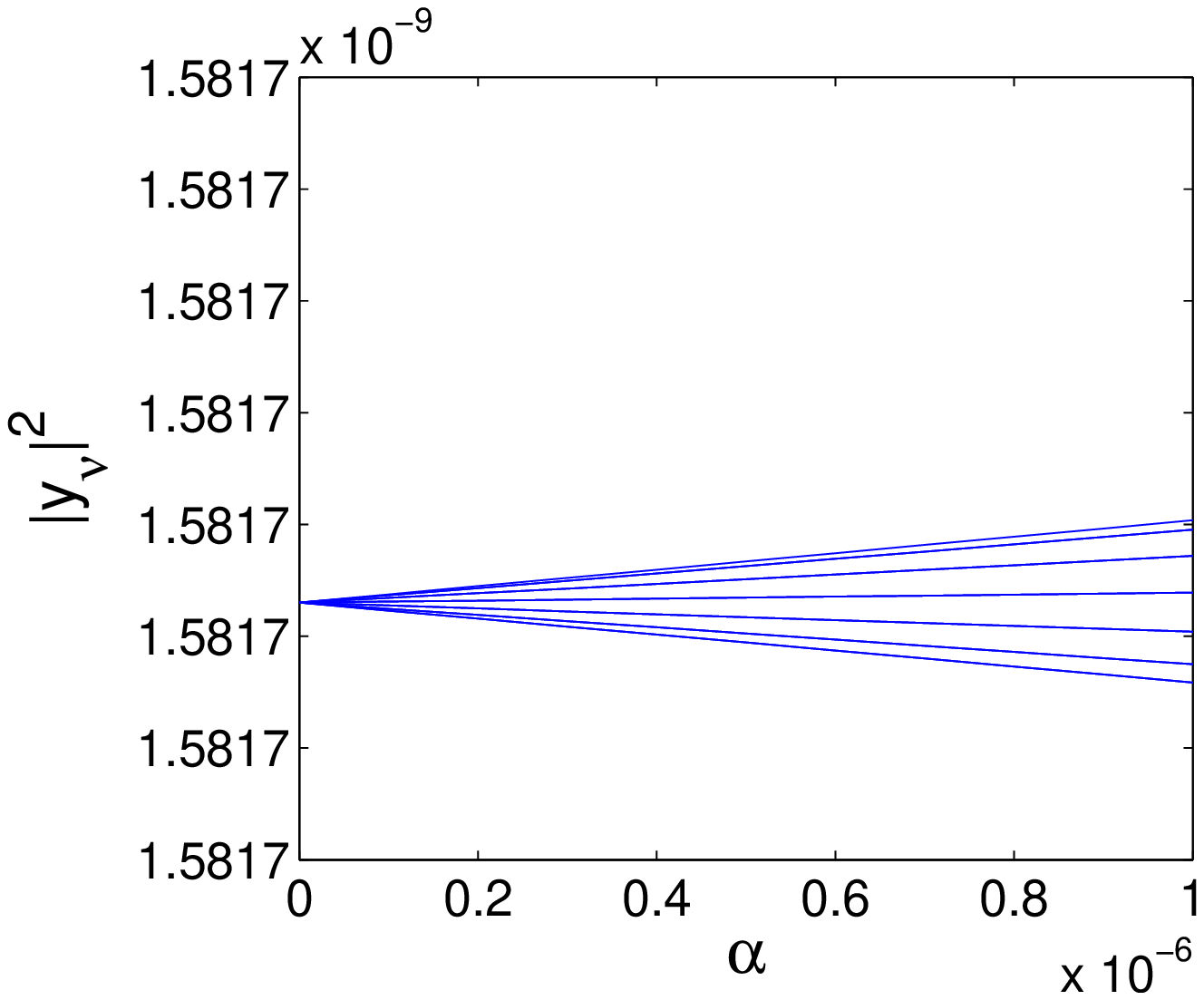}}
\subfloat[$Neutron$]{\includegraphics[width=2in]{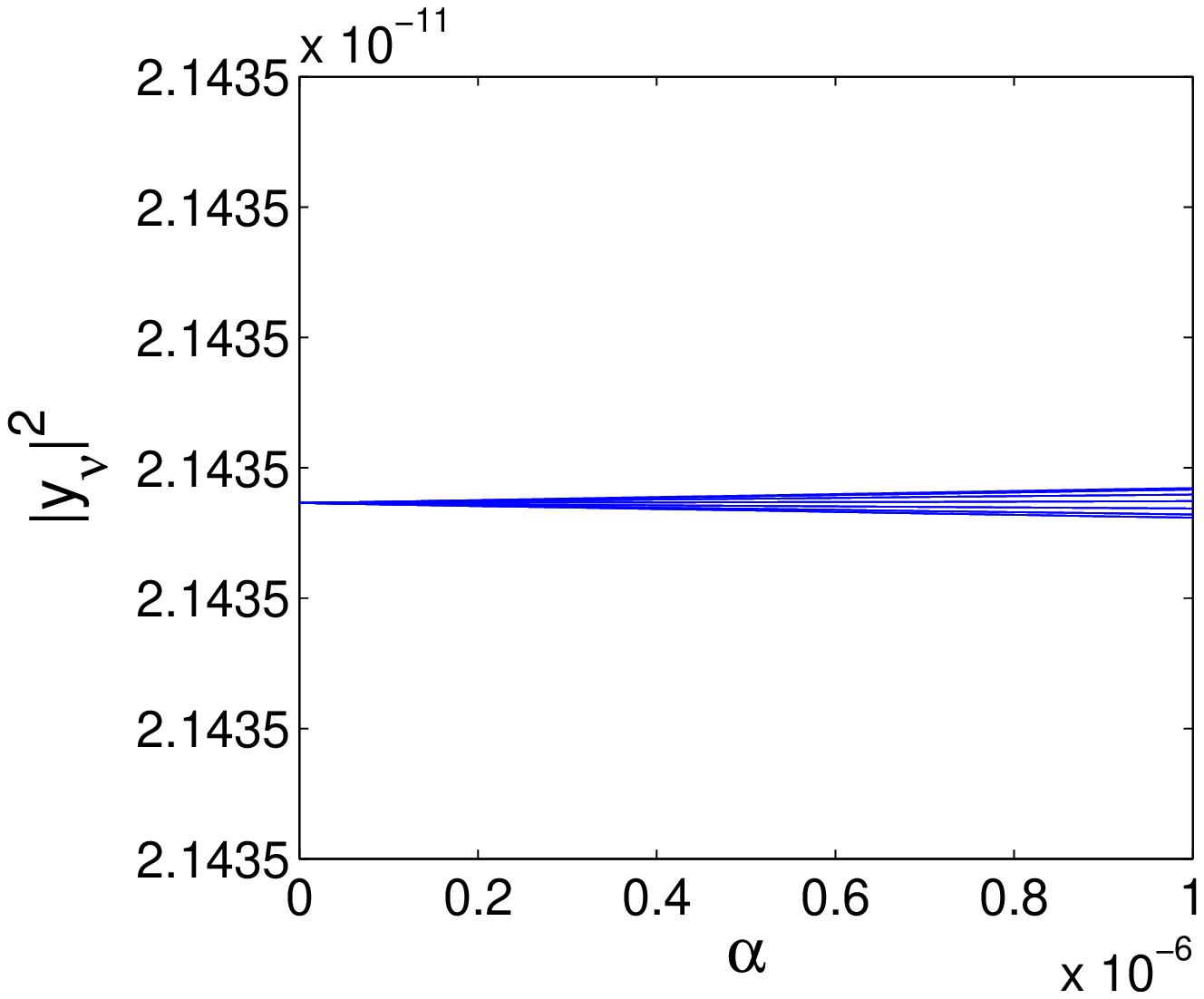}}
\caption{The $|y_\nu|^2$ for different particles, the differences between different $|y_\nu|^2$ determine continuity of the energy decrease quickly when the mass of particles decrease, where we have chosen $\alpha=0.800$.}\label{fig:2}
\end{figure*}

On the other hand, we also find the effect of the mass on the continuity of the energy. Figure \ref{fig:2} illustrates the difference of different $|y_\nu|^2$ vs. particle mass for three different particles. The figure illustrates that the differences between different $|y_\nu|^2$ become large when the mass of the particle becomes small. When the particle is a neutron, the difference between different $|y_\nu|^2$ is very small. In other words, the continuity of the energy will become strong when the particle mass decreases; conversely, all the energy will be asymptotically close to the case in classical quantum mechanics when the particle mass becomes sufficiently large, which coincides with the results in the experiments \cite{12,14}.\par
\begin{figure}[!h]
\centering
\includegraphics[width=5in]{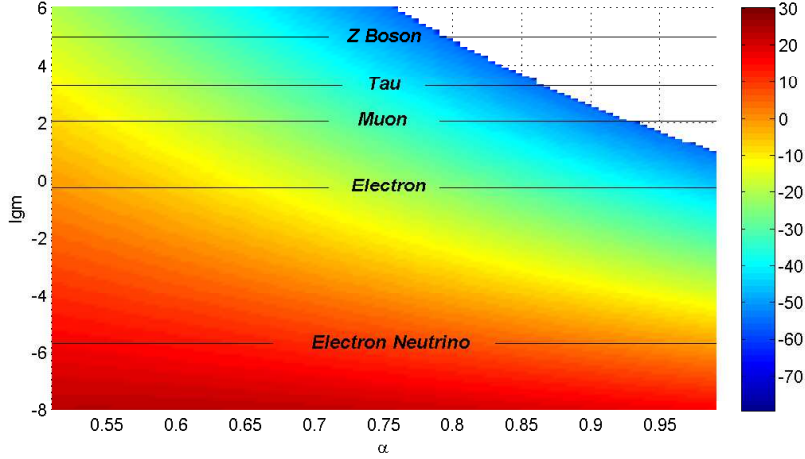}
\caption{The magnitude of the variances of $|y_\nu|^2$ for different masses and L\'{e}vy path dimensions, where we choose $\xi=10^{-6}m$, $D_{2\alpha}={c^{2-2\alpha}}/{(2\alpha m^{2\alpha-1})}$. For $\alpha\in(0.5,1)$, the red colors denote that the variances of $|y_\nu|^2$ are large, and the parameters have sufficient degrees of freedom; thus, the energies of the bound states are continuous. In addition, the blue colors and the white color indicate that the variances of $|y_\nu|^2$ are small, and the parameters have few degrees of freedom; thus, the energies of bound states are discrete (The white color indicates that the variances of $|y_\nu|^2$ are less than $10^{-80}$).\label{fig:3}}
\end{figure}

As another main conclusion, the continuity of the energy will become strong when (1) the mass of the particle becomes small and/or (2) the
dimension of the L\'{e}vy path changes from integer to non-integer. Figure \ref{fig:3} illustrates the changes between $|y_\nu|^2$ at the point $\xi=10^{-6}m$ for different $\alpha$ and different masses in the Earth gravitational field. For ultralight particles, such as the electron neutrino, the energies of bound states are always continuous in fractional dimensions. However, for the electron, the energies of bound states are continuous only in the region far from $1$ (classical situation). In addition, for particles with large masses, such as muon, tau, and Z boson, the continuity of the energy is obviously weak. Since the experiments involved the observation of the energy levels of particles in the Earth's gravitational field \cite{12}, it is worth noting that the results indicate that the conspicuous continuity of energy is evidence of the existence of ultralight particles, which may improve methods of detecting ultralight particles.\par

\subsection{The minimal length}\label{sec:3}
To study influence to the above phenomena of phase transitions and BICs given by minimal length, we consider a particle moving in the Earth's gravitational field with the minimal length in fractional $2\alpha$ dimensions. Here, as discussed above, we assume that the mass of the particle $m<10^{-26}kg$ which cause an obvious characteristic of ultralight particles. The start point of the study is the generalized uncertainty principle (GUP) \cite{23} in one dimension,
\begin{equation*}
\Delta x\Delta p\geq\frac{\hbar}{2}\{1+3\beta[(\Delta p)^2+(Ep)^2]\},
\end{equation*}
which equals the modified Heisenberg algebra \cite{22}:
\begin{equation}
[x,p]=i\hbar(1+3\beta p^2),
\end{equation}
where $\beta=({l_pl}^2)/(2\hbar^2)$ and ${l_pl}^2=(G\hbar)/{c^3}$ is the square of the Planck length.\par
In the position representation, $x$ and $p$ can be defined as
\begin{equation}\label{eq:GUP}
x=x_0,p=p_0(1+\beta {p_0}^2),
\end{equation}
where $x_0$ and $p_0$ satisfy the canonical commutation relation $[x,p]=i\hbar$ \cite{17}.\par
By the fractional binomial formula $(1+x)^\alpha=\sum_{k=0}^{+\infty}{\alpha \choose k}x^k$ and plugging Eq. \eqref{eq:GUP} into the Hamiltonian Eq. \eqref{eq:ham}, we have
\begin{equation}\label{eq:Hamiltonian}
H=H_0+H_1,
\end{equation}
where $H_0=D_{2\alpha}p^{2\alpha}+V(x)$ is a usual Hamiltonian with fractional dimension and $H_1=D_{2\alpha}[2\alpha\beta p^{2(\alpha+1)}+O(\beta^2)]$ is a perturbation, which is regarded as the universality of quantum gravity corrections \cite{17}. Next, we will study the influence on the phase transition of the energy and wave functions, and BICs in the gravitational field due to $H_1$.\par

By the fractional corresponding relation Eq. \eqref{eq:corresponding}, Eq. \eqref{eq:Hamiltonian} becomes
\begin{equation}
H=D_{2\alpha}[(-i\hbar)^{2\alpha}T_{2\alpha}+2\alpha\beta (-i\hbar)^{2(\alpha+1)}T_{2(\alpha+1)}+O(\beta^2)]+V(x),
\end{equation}
Without loss of generality, we choose $T_{2\alpha}$ as the R-L fractional derivative operator and $V(z)$ as the gravitational field defined by Eq. \eqref{eq:gravitational field}, and we neglect terms of order $\beta^2$ and higher,

Then, the fractional Schr\"{o}dinger equation in the presence of a minimal length is
\begin{equation}\label{eq:FSEML}
D_{2\alpha}[(-i\hbar)^{2\alpha}{}_{-\infty}^R\!D_z^{2\alpha}\varphi(z)+2\alpha\beta (-i\hbar)^{2(\alpha+1)}{}_{-\infty}^R\!D_z^{2(\alpha+1)}\varphi(z)]+V(z)\varphi(z)=E\varphi(z).
\end{equation}\par
We can see that there are two fractional derivative operators in Eq. \eqref{eq:FSEML}, and they make Eq. \eqref{eq:FSEML} difficult to solve. Thus, we introduce the auxiliary function $\varphi(z)=(1-2\beta\alpha p^2)\phi(z)$ mentioned in \cite{18} to reduce Eq. \eqref{eq:FSEML} to only having one fractional derivative operator.\par
Letting $A=E-2\alpha\beta D_{2\alpha}^{1+1/\alpha}{E^{0}}^{1+1/\alpha}$, $B=mg[2\alpha\beta D_{2\alpha}^{1+1/\alpha}(E^{0}/\alpha+{E^{0}}^{1/\alpha})-1]$, $\xi=({Az+B})/[{D_{2\alpha}(-i\hbar)^{2\alpha}}]$, $K=\{[{D_{2\alpha}(-i\hbar)^{2\alpha}}]/{A}\}^{2\alpha}$, and $\psi[\xi(z)]=\phi(z)$,
Eq. \eqref{eq:FSEML} becomes approximatively to (see Appendix \ref{app-c} for related mathematical deductions)

\begin{equation}\label{eq:FE1}
{}_{-\infty}^R\!D_\xi^{2\alpha}\psi(\xi)-K\xi\psi(\xi)=0.
\end{equation}\par
The solutions of Eq. \eqref{eq:FE1} (see Appendix \ref{app-a} for a description of this method and related mathematical deductions) are
\begin{equation}
\psi(\xi)=\sum C_\nu y_\nu(\xi),
\end{equation}
where
\begin{equation*}
y_\nu(\xi)=\varepsilon_\nu\int_0^{+\infty}exp\{\varepsilon_\nu zt-\frac{t^{2\alpha+1}}{({D_{2\alpha}(-i\hbar)^{2\alpha}}/{A})^{2\alpha}(2\alpha+1)}\}dt,
\end{equation*}
$\varepsilon_\nu^{2\alpha+1}=1$ and $\{C_\nu\}$ is a sequence of arbitrary coefficients that satisfy $\sum C_\nu=0$.\par
It is shown that the minimal length does not change the value of $\varepsilon_\nu$ as presented in Fig. \ref{fig:1}(a)-(c) for different $\alpha$. Thus, the minimal length does not change the number of free undetermined parameters. From the discussion in the above section, we can find the following: If $\alpha\in(0.5,1)$, the energies of the bound states are continuous, and every energy level $E$ has degeneracy. This is strongly related to the dimension $2\alpha$ of the L\'{e}vy path. That is the nonzero minimum length does not change the energy in the fractional case; then an energy shift does not occur, which is a common phenomenon in the presence of a minimal length. Thus, the phase transitions and BICs still exist. If $\alpha=1$, the nonzero minimum length will give a correction to the energy; there is an energy shift \cite{48,49}. \par

\section{Conclusion}
We study a particle moving in the Earth's gravitational field in the circumstance of the fractional L\'{e}vy path. We obtain analytical expressions for the wave functions and energy, and we find the phase transitions of wave functions and energy: the energy changes from discrete to continuous and wave functions change from non-degenerate to degenerate when dimension of L\'{e}vy path becomes from integer to non-integer. The particle's mass and the dimensions of the L\'{e}vy path produce and influence the phase transitions: the continuity of energy becomes strong when (1) the dimension of the L\'{e}vy path changes from integer to non-integer and/or (2) the mass of the particle becomes small. Moreover, our results remain valid under the minimal length because energy shift does not occur. Our results indicate two main conclusions. One conclusion is that the phase transitions can provide a simpler method to produce BICs. Since there is evidence that the BICs phenomenon can be observed in some experiments \cite{69}, in the best case scenario, it may provide an experimental method to verify the successful preparation of a fractional quantum system. The other conclusion is that the phase transitions offer a conspicuous phenomenon to verity the existence of ultralight particles \cite{69} such as the electron neutrino and the positive electron. Since the experiments involve the observation of the energy levels of particles in the Earth's field \cite{12}, we believe this may improve methods of detecting ultralight particles.
\section{Acknowledgements}
We also thank professor Ma Hong for his thoughtful suggestions about fractional quantum mechanics.
\appendix
\section{Method of solving fractional differential equations}\label{app-a}
We present a new method of solving fractional differential equations of the form
\begin{equation}\label{eq:GFSE-appendix}
{}_{-\infty}^R\!D_\xi^{2\alpha}\varphi(\xi)-K\xi\varphi(\xi)=0, (\alpha\in(1,2], K>0).
\end{equation}\par
We introduce an auxiliary function $y_\nu$ as follows:
\begin{equation}\label{eq:aux0-appendix}
y_\nu(\xi)=\varepsilon_\nu\int_0^{+\infty}\exp[\varepsilon_\nu \xi t-\frac{t^{2\alpha+1}}{K(2\alpha+1)}]\mathrm{d}t,
\end{equation}
where $\varepsilon_\nu^{2\alpha+1}=1$.\par
Note that
\begin{equation}\label{eq:g-middle-appendix}
\begin{split}
{}_{-\infty}^R\!D_\xi^{2\alpha}y_\nu(\xi)=
&\varepsilon_\nu K\int_0^{+\infty}\exp[\varepsilon_\nu \xi t-\frac{t^{2\alpha+1}}{K(2\alpha+1)}]\mathrm{d}[{\frac{\varepsilon_\nu^{2\alpha}}{K(2\alpha+1)}t^{2\alpha+1}}].
\end{split}
\end{equation}\par
Substituting Eq. \eqref{eq:g-middle-appendix} and $\varepsilon_\nu^{2\alpha+1}=1$ into ${}_{-\infty}^R\!D_\xi^{2\alpha}y_\nu(\xi)-K\xi y_\nu(\xi)$, we have
\begin{equation*}
\begin{split}
{}_{-\infty}^R\!D_\xi^{2\alpha}y_\nu(\xi)-K\xi y_\nu(\xi)
& =\varepsilon_\nu K \int_0^{+\infty}\exp[\varepsilon_\nu \xi t-\frac{t^{2\alpha+1}}{K(2\alpha+1)}]\mathrm{d}[{\frac{\varepsilon_\nu^{2\alpha}}{K(2\alpha+1)}t^{2\alpha+1}}]\\
& -\varepsilon_\nu K\xi \int_0^{+\infty}\exp[\varepsilon_\nu \xi t-\frac{t^{2\alpha+1}}{K(2\alpha+1)}]\mathrm{d}t\\
& =- K \int_0^{+\infty}\exp[\varepsilon_\nu \xi t-\frac{t^{2\alpha+1}}{K(2\alpha+1)}]\mathrm{d}[\varepsilon_\nu \xi t-\frac{t^{2\alpha+1}}{K(2\alpha+1)}]\\
& =- K \exp[\varepsilon_\nu \xi t-\frac{t^{2\alpha+1}}{K(2\alpha+1)}]|_0^{+\infty}.
\end{split}
\end{equation*}\par
Since Laskin showed that $2\alpha$ must be in $(1,2]$ in quantum mechanics \cite{6,7}, we have $\varepsilon_\nu \xi t-{t^{2\alpha+1}}/{K(2\alpha+1)}\rightarrow -\infty$ as $t\rightarrow +\infty$. Then, we can obtain
\begin{equation}\label{eq:g-middle2-appendix}
{}_{-\infty}^R\!D_\xi^{2\alpha}y_\nu(\xi)-K\xi y_\nu(\xi)=K.
\end{equation}\par
Assuming that $\{C_\nu\}$ is a sequence of arbitrary coefficients that satisfies $\sum C_\nu=0$, by $\sum C_\nu=0$ and Eq. \eqref{eq:g-middle2-appendix}, for an energy level $E$, we can find that the solutions of Eq. \eqref{eq:GFSE-appendix} are
\begin{equation}\label{eq:g-solution-appendix}
\varphi(\xi)=\sum C_\nu y_\nu(\xi).
\end{equation}\par
\section{Minimum of numerator of $\frac{2n+m}{m}$}\label{app-b}
We will discuss this in three different cases.\par
Case 1\par
Assume that $m$ is even and that $n$ is odd. Thus, we can let $m=2m_1$, where $m_1$ is an integer. Since the greatest common factor of $m,n$ is $(m,n)=1$, we have $(m_1,n)=1$.\par

Then
\begin{equation*}
\frac{2n+m}{m}=\frac{n+m_1}{m_1}.
\end{equation*}\par
Since $(m_1,n)=1$ and $(m_1,m_1)=m_1$, we obtain $(n+m_1,m_1)=1$. For $\alpha$ in the interval $(0.5,1]$, when $\frac{n}{m}=\frac{3}{4}$, the numerator of $\frac{2n+m}{m}$ obtains its minimum of $5$.\par
Case 2\par
Assume that $m$ is odd, and that $n$ is even. Then $(2n,m)=1$, and $(2n+m,m)=1$. For $\alpha$ in the interval $(0.5,1]$, when $\frac{n}{m}=\frac{2}{3}$, the numerator of $\frac{2n+m}{m}$ obtains its minimum of $7$.\par
Case 3\par
Assume that $m$ is odd, and that $n$ is odd. Since $(m,n)=1$, $(2n+m,m)=1$. For $\alpha$ in the interval $(0.5,1]$, when $\frac{n}{m}=\frac{3}{5}$, the numerator of $\frac{2n+m}{m}$ obtain its minimum of $11$.\par
The above three cases cover all irreducible fractions; thus, the minimum of the numerator of $\frac{2n+m}{m}$ is $5$.
\section{Derivation of Eq. \eqref{eq:FE1}}\label{app-c}
Plugging $\varphi(z)=(1-2\beta\alpha p^2)\phi(z)$ into Eq. \eqref{eq:FSEML} and neglecting terms of order $\beta^2$ and higher, we have
\begin{equation}\label{eq:aux1}
D_{2\alpha}(-i\hbar)^{2\alpha}{}_{-\infty}^R\!D_z^{2\alpha}\phi(z)+2[V(z)-E]\beta\alpha\hbar^2\frac{\partial^2}{\partial z^2}\phi(z)+[V(z)-E]\phi(z)=0.
\end{equation}\par
For further simplicity, we consider the fractional Schr\"{o}dinger equation as follows:
\begin{equation*}
[D_{2\alpha}p^{2\alpha}+V(z)]\phi(z)=E\phi(z),
\end{equation*}
we know that $p^{2\alpha}=D_{2\alpha}^{-1}[E^0-V(z)]$, i.e., $p^2=\{D_{2\alpha}^{-1}[E^0-V(z)]\}^{1/\alpha}$, where $E^{0}$ is the eigenvalue of $H_0$. Thus, Eq. \eqref{eq:aux1} becomes
\begin{equation}\label{eq:aux1.2}
\begin{split}
&D_{2\alpha}(-i\hbar)^{2\alpha}{}_{-\infty}^R\!D_z^{2\alpha}\phi(z)+[V(z)-E]\phi(z)\\
&+2\alpha\beta[E-V(z)]\{D_{2\alpha}^{-1}[E^0-V(z)]\}^{1/\alpha}\phi(z)=0.
\end{split}
\end{equation}\par
In the region $z>0$, $V(z)=mgz$. Using the fractional binomial formula, we obtain
\begin{equation}\label{eq:aux-b}
{(E^0-mgz)}^{1/\alpha}=\sum_{k=0}^{+\infty}{1/\alpha \choose k}{E^0}^{\frac{1}{\alpha}}{(-\frac{mgz}{E^0})}^k.
\end{equation}\par
As discussed in Section \ref{sec:2b}, for large-mass particle, the phase transitions and BICs are not conspicuous; thus, here, we only consider small-mass particle. Since $D_{2\alpha}$ depends on $m$, extending the terms in Eq. \eqref{eq:aux1.2} containing $\beta m$ to a first-order approximation after we substitute Eq. \eqref{eq:aux-b} into it, we have
\begin{equation}\label{eq:aux1.5}
\begin{split}
&D_{2\alpha}(-i\hbar)^{2\alpha}{}_{-\infty}^R\!D_z^{2\alpha}\phi(z)+2\alpha\beta D_{2\alpha}^{1+1/\alpha}(E{E^0}^{1/\alpha}\\
&-\frac{1}{\alpha}E{E^0}^{1/\alpha-1}mgz-{E^0}^{1/\alpha}mgz)\phi(z)+(mgz-E)\phi(z)=0.
\end{split}
\end{equation}\par
Letting $A=E-2\alpha\beta D_{2\alpha}^{1+1/\alpha}E{E^0}^{1/\alpha}$ and $B=mg[2\alpha\beta D_{2\alpha}^{1+1/\alpha}(\frac{1}{\alpha}E{E^0}^{1/\alpha-1}+{E^0}^{1/\alpha})-1]$, where $E^0$ is the eigenvalue of $H_0$, and Eq. \eqref{eq:aux1.5} can be simplified to
\begin{equation}\label{eq:aux2}
D_{2\alpha}(-i\hbar)^{2\alpha}{}_{-\infty}^R\!D_z^{2\alpha}\phi(z)-(Az+B)\phi(z)=0.
\end{equation}\par
To use the method of solving fractional differential equations that we built in Appendix \ref{app-a}, we need to let $\xi=(Az+B)/[D_{2\alpha}(-i\hbar)^{2\alpha}]$, $K=\{[{D_{2\alpha}(-i\hbar)^{2\alpha}}]/{A}\}^{2\alpha}$, and $\psi[\xi(z)]=\phi(z)$. Then, using the fractional chain rule \cite{25}, Eq. \eqref{eq:aux2} becomes
\begin{equation}\label{eq:FE1-appC}
{}_{-\infty}^R\!D_\xi^{2\alpha}\psi(\xi)-K\xi\psi(\xi)=0.
\end{equation}\par
Obviously, Eq. \eqref{eq:FE1-appC} can be solved by the method that we built in Appendix \ref{app-a}.
\section{Different effects of different fractional derivatives}\label{app-d}
1. The core of Gr\"{u}nwald-Letnikov fractional derivative is generalizing the limit of the concept of difference
\begin{equation*}
\frac{d^n}{\mathrm{d}x^n}f(x)
 = \lim\limits_{h\rightarrow0}\frac{1}{h^n}\sum_{k=0}^n(-1)^k\left(
                                                                  \begin{array}{c}
                                                                    n \\
                                                                    k \\
                                                                  \end{array}
                                                                \right) f(x-kh)\\,
\end{equation*}
to fractional cases:

\begin{equation*}
{}_{a}^G\!D_x^\alpha f(x)
 = \lim\limits_{
                    h\rightarrow0,nh=t-a
}\frac{1}{h^\alpha}\sum_{k=0}^n\left[
                                                                  \begin{array}{c}
                                                                    \alpha \\
                                                                    k \\
                                                                  \end{array}
                                                                \right] f(x-kh)\\,
\end{equation*}
where $\left(
                                                                  \begin{array}{c}
                                                                    n \\
                                                                    k \\
                                                                  \end{array}
                                                                \right)$ are binomial coefficients, and $\left[
                                                                  \begin{array}{c}
                                                                    \alpha \\
                                                                    k \\
                                                                  \end{array}
                                                                \right]=\frac{\alpha(\alpha-1)\ldots(\alpha-k+1)}{k!}$ are generalized binomial coefficients.\par
The advantage of difference is that can easily realize numerical calculation. Compared with classical derivative, it needs lofty mathematical skill to calculate fractional derivative to complex functions and composite functions. And also compared with other fractional derivatives, Gr\"{u}nwald-Letnikov fractional derivative has a complex expression:
\begin{equation}\label{eq:GL}
\begin{split}
{}_{a}^G\!D_x^\alpha f(x,t)
&=\sum^m_{k=0}\frac{\frac{\partial^k f(a,t)}{\partial x^k}(x-a)^{-\alpha+k}}{\Gamma(-\alpha+k+1)}\\
&+\frac{1}{\Gamma(-\alpha+m+1)}\int_a^x(x-u)^{m-\alpha}\frac{\partial^{m+1} f(u,t)}{\partial x^{m+1}}\mathrm{d}u.
\end{split}
\end{equation}
Thus, Gr\"{u}nwald-Letnikov fractional derivative is usually considered to use in the cases needing numerical calculation.\par
2. Compared Gr\"{u}nwald-Letnikov fractional derivative Eq. \eqref{eq:GL} and Riemann-Liouville fractional derivative
\begin{equation*}
{}_{a}^R\!D_x^\alpha f(x,t)=\frac{1}{\Gamma(m-\alpha)}\frac{\partial^m}{\partial x^m}\int_a^x(x-u)^{m-\alpha-1}f(u,t)\mathrm{d}u.
\end{equation*}
Note that Gr\"{u}nwald-Letnikov fractional derivative needs to calculate $m$-order derivative of the function $f$ and it contains a sum of the infinite series, so Riemann-Liouville fractional derivative has a wide domain, and can calculate the fractional derivative without considering convergence in the sum of the infinite series.\par
3. The initial conditions for equations with Caputo fractional derivative used the same form as for integer-order differential equations. And also Caputo fractional derivative of a constant is $0$. The two advantages make Caputo fractional derivative with more physical significance to describe problems; however, under the conditions of this paper, that is using the lower bound of integral with $-\infty$, Riemann-Liouville fractional derivative and Caputo fractional derivative is equivalent. Thus, Riemann-Liouville fractional derivative also has the above two advantages of Caputo fractional derivative.\par
4. We find when the problem with some barrier that particle can not across, using the fractional derivative based on generalized functions to build the fractional Schr\"{o}dinger equation, the equation already contains some information of the boundary conditions.\par
Assume at the point $x=0$, the potential is $V(0)=+\infty$. If a particle moves in the region $x>0$, it can pass through the region $x>0$. Without loss of generality, we assume $V=0$ for $x>0$.\par
We also assume the wave function is $\varphi$, then the bound condition at the point $x=0$ is
\begin{equation}\label{eq:temp5}
\varphi|_{x=0}=0.
\end{equation}\par
By separation of variables, we only need solve the time-independent wave function $\varphi(x)$. If we use the fractional derivative based on generalized functions ${}_{-\infty}\!\tilde{D}_x^{2\alpha}$ to describe states of the particle in the region $x>0$, then the fractional Schr\"{o}dinger equation is
\begin{equation}\label{eq:FSEBtemp}
D_{2\alpha} (-i\hbar)^{2\alpha} {}_{-\infty}\!\tilde{D}_x^{2\alpha}\varphi(x)=E\varphi(x).
\end{equation}\par
By the generalized function $\Phi_{-\alpha}(x)$ of ${}_{-\infty}\!\tilde{D}_x^{2\alpha}$ is
\begin{equation*}
\Phi_{-\alpha}(x)=
\begin{cases}
\frac{x^{-\alpha-1}}{\Gamma(-\alpha)} & \text{$x>0$};\\
                                  0 & \text{$x\leq0$},
\end{cases}
\end{equation*}
we have
\begin{equation}\label{eq:temp4}
{}_{-\infty}\!\tilde{D}_x^\alpha f(x)=f(x)*\Phi_{-\alpha}(x)=0,x\leq0.
\end{equation}\par
Substituting Eq. \eqref{eq:temp4} into Eq. \eqref{eq:FSEBtemp}, we obtain
\begin{equation*}
\varphi(x)=\frac{1}{E}[D_{2\alpha} (-i\hbar)^{2\alpha} ]{}_{-\infty}\!\tilde{D}_x^{2\alpha}\varphi(x)=0,x\leq0,
\end{equation*}
That is $\varphi(x)=0$ when $x=0$. Thus, we obtain the bound condition Eq. \eqref{eq:temp5} from the fractional Schr\"{o}dinger equation Eq. \eqref{eq:FSEBtemp} directly. This means when we deal with some problems with some barrier that particle can not across, the equation already contains some information of the boundary conditions if we use the fractional derivative based on generalized functions.
\bibliographystyle{unsrt}
\bibliography{beginning3reference}
\end{document}